\documentclass[twocolumn,showpacs,amsmath,amssymb,prl,superscriptaddress]{revtex4-1}

 \usepackage{graphicx}
 \usepackage{amsmath}
 \usepackage{amsfonts}
 \usepackage{amssymb}
 \usepackage{pstricks}
 \usepackage{psfrag}
 \usepackage{epsfig}
 \usepackage[utf8]{inputenc}

\usepackage[textsize=footnotesize, backgroundcolor=yellow!10, linecolor=gray!35]{todonotes}

\begin{document}

\title{Crawling in a fluid}

\author{Alexander Farutin}
\email{alexandr.farutin@univ-grenoble-alpes.fr}
\author{Jocelyn \'Etienne}
\author{Chaouqi Misbah}
\author{Pierre Recho}
\affiliation{Univ. Grenoble Alpes, CNRS, LIPhy, F-38000 Grenoble, France}

\date{Received: \today / Revised version: (date)}

\begin{abstract}
There is increasing evidence  that mammalian cells not only crawl on  substrates but can
also swim in fluids. To elucidate the mechanisms of the onset of
motility of cells in suspension, a model which couples actin and myosin kinetics  to
fluid flow is proposed and solved  for a spherical shape. The swimming speed is extracted in terms of key parameters.
We analytically find super- and subcritical bifurcations from a non-motile to a motile
state and also spontaneous polarity oscillations that arise from a Hopf bifurcation. Relaxing
the spherical assumption, the obtained shapes show appealing trends.
\end{abstract}


\maketitle

\paragraph{Introduction}

Cell motility plays an essential role during embryogenesis, tissue renewal and repair, response of the immune system to infection as well as pathological processes such as cancer cell migration. 
It is also important from an engineering point of view to conceive biomimetic robots.

A longstanding dogma in biology is that mammalian cells need mechanical interaction with
a solid substrate in order to move forward.
The commonly adopted idea about the source of motion is that actin polymerization and actin contraction assisted by molecular motors (myosin) create a tangential flow of the cell cortex,
the retrograde flow. Thrust is then achieved by the transmission of momentum 
to the outer environment by a dynamic adhesive coupling through
transmembrane proteins such as integrins \cite{mogilner2009mathematics}. Several modes of such crawling motion are distinguished in the literature, most of which involve ample shape changes \cite{petrie2015fibroblasts, zhu2016comparison,aroush2017actin,paluch2013role}
but crawling in a shape-preserving manner is also possible, as notably observed for fish keratocytes \cite{Lee1957,Rubinstein+Verkhovsky-Mogilner.2009.1,Tjhung+Cates.2015.1,ziebert2011model}.
A major step forward has been achieved by identifying that leukocytes  can
migrate in an extracellular matrix without the assistance of integrins
\cite{lammermann2008rapid}, raising the important question about the role
of specific adhesion in crawling. Indeed, it is reported that crawling can be mediated by friction only \cite{Bergert2015} when the cell is confined by a 3D environment. Increasing attention has been recently paid to the understanding of such adhesion-independent crawling \cite{paluch_review,Renkawitz2009,Hawkins2009}.

However, it is also reported that some cells can swim in an unconfined fluid 
\cite{barry2010dictyostelium,arroyo2012reverse,Theodoly}. Swimming relies on periodic 
distortions of the cell surface which generate a reaction force from the ambient fluid
\cite{lauga2009hydrodynamics}. Beside the special case of ciliated squirmers 
\cite{stone1996propulsion}, the mainstream notion of swimming relies on
ample shape changes, the ones of flagellas \cite{lauga2009hydrodynamics} or of 
the cell body \cite{brennen1977fluid,arroyo2012reverse,fackler2008cell,Farutin2013,campbell2017computational}.

In this letter, we show that swimming may be generically possible also for mammalian cells through the same growth and contraction driven cortical flows operating during crawling and in the absence shape changes.  Mechanical interaction with a substrate or extracellular matrix, and ample shape changes are therefore not necessary for crawling. This actuation mechanism necessitates spatial symmetry breaking in order to
generate directed motion, in the same way as in the case of crawling on a
solid substrate \cite{verkhovsky1999self, recho2013contraction,barnhart2015balance}.  
Recently, symmetry breaking in cortex dynamics and the resulting cell shapes have been analyzed for quasi-spherical cells \cite{Voituriez2016} but the direct relation between the symmetry breaking and the swimming motion has not
been described. By coupling the acto-myosin dynamics to the internal, external, and cortex flow, we provide analytically the functional dependence of the swimming speed on the relevant parameters.

A second finding is the identification of a rich panel of instabilities leading to cell polarization.
Depending on three key parameters of the model (actin stiffness, rate of actin turn-over, myosin contractility) we find either a supercritical bifurcation,  leading to a
continuous transition between static and motile configurations, or a subcritical bifurcation implying a metastable coexistence between the static and motile configurations for a finite range of parameters.
For a small enough actin turn-over rate, the system exhibits a Hopf bifurcation resulting in a permanent oscillation of the polarization.
Such behaviour is reminiscent of recent experimental evidences of actomyosin oscillations \cite{lavi2016deterministic, godeau2016cyclic}. Finally, we show the shapes of polarized cells if the fixed shape assumption is relaxed.
The shapes are obtained in a quasi-spherical limit and show reasonable trends.

\paragraph{Model}
We consider a neutrally buoyant cell in a fluid environment.
Swimming is achieved by transmission of shear stress by the cortex flow to the surrounding fluid.
We leave the question of the exact nature of the transmission mechanism and its efficiency outside the scope of this study, and assume perfect no-slip conditions between the cortex and the fluids inside and outside the cell.

We first focus on a fixed spherical shape and show how spontaneous polarization of acto-myosin dynamics and the resulting cortex flow can drive the deformation-independent swimming of the cell.
The cortex of the cell is modeled as a compressible two-dimensional (2D) Newtonian fluid along the surface of the sphere.
The surface density of actin meshwork is denoted $c^a.$
The distribution of myosin that crosslinks actin is characterized by the concentration $c^\mu.$
The velocity field (in the laboratory frame) of the cortex is denoted as $\boldsymbol u^c$. 
The surface strain rate tensor reads $\mathsf\epsilon^s=\boldsymbol\nabla^s\otimes \boldsymbol u^c\boldsymbol\cdot \mathsf I^s+\mathsf I^s \boldsymbol\cdot (\boldsymbol\nabla^s\otimes \boldsymbol u^c)^T$ where $\mathsf I^s$ is the surface projection operator and $\nabla^s\equiv\mathsf I^s\boldsymbol\cdot\boldsymbol\nabla$ is the surface gradient. 
In the framework of the active gel theory \cite{Kruse+.2004.1}, we use the following constitutive equation for the surface stress
\begin{equation}
\label{surfacestress}
\mathsf \sigma^s=\eta_s\epsilon^s+\eta_b\mathsf I^s (\boldsymbol \nabla^s\boldsymbol\cdot \boldsymbol u^c)+\mathsf I^s(\chi c^\mu-\alpha c^a),
\end{equation}
where $\eta_s$ and $\eta_b$ are 2D shear and bulk viscosities, $\chi$ is the myosin-induced contractility and $\alpha$ is the 2D bulk modulus of the actin meshwork. 

The  surface forces are calculated as $\boldsymbol f=\boldsymbol\nabla^s\boldsymbol\cdot\mathsf\sigma^s-f^n\boldsymbol n$, where $\boldsymbol n$ is the outward normal and $f^n$ is
a Lagrange multiplier that enforces the fixed shape \cite{Farutin2012}.
It satisfies the
zero total force condition $\oint f^n\boldsymbol n dA=0$, where $dA$ is the area element on the boundary of the cell.
%
The forces $\boldsymbol f$ are balanced by viscous stresses of the cytoplasm (the fluid occupying the cell interior) and the suspending fluid at the cell boundary.
Both are considered incompressible Newtonian fluids of viscosities $\eta_{in}$ and $\eta_{out}$, respectively.
With this assumption and at zero Reynolds number, 3D fluid dynamics is governed by Stokes equations, which can be combined into a boundary integral equation \cite{Pozrikidis1992}
\begin{align}\label{BIE}
&\frac{\eta_{in}+\eta_{out}}{2}u_i^c(\boldsymbol x)=\oint\limits G_{ij}(\boldsymbol x,\boldsymbol x')f_j(\boldsymbol x')dA(\boldsymbol x')\\
+&(\eta_{out}-\eta_{in})\oint\limits K_{ijk}(\boldsymbol x,\boldsymbol x') u_j^c(\boldsymbol x') n_k (\boldsymbol x') dA(\boldsymbol x')\notag
\end{align}
using the boundary conditions at infinity, as well as stress balance and continuity of velocity at the cell surface. 
Here $\boldsymbol x$ and $\boldsymbol x'$ are points on the cell surface.
The Green's kernels $G_{ij}$ and $K_{ijk}$ are listed in \cite{SInfo}.

The swimming speed is defined as the volume-averaged velocity of the cytoplasm, which can be converted ino a surface integral $\boldsymbol v_s = (\frac{4}{3}\pi R^3)^{-1} \oint\boldsymbol x(\boldsymbol u^c\boldsymbol\cdot\boldsymbol n) dA$, where $R$ is the cell radius. The fixed shape assumption implies that the cortex velocity must be tangential to the cell surface in the comoving frame: $(\boldsymbol u^c-\boldsymbol v_s)\boldsymbol\cdot\boldsymbol n=0$.
This condition fixes $ f^n$.

To close the description we express the conservation equations on the actin and myosin fields $c^a$ and $c^\mu$:
\begin{eqnarray}
\label{advection1}
&\dot c^a+\boldsymbol\nabla^s\boldsymbol\cdot\left[c^a (\boldsymbol u^c-\boldsymbol v_s)\right]=\beta (c^a_0-c^a)\\
\label{advection2}
&\dot c^\mu+\boldsymbol\nabla^s\boldsymbol\cdot\left[c^\mu (\boldsymbol u^c-\boldsymbol v_s)\right]=D^\mu\Delta^sc^\mu,
\end{eqnarray}
where dot denotes time derivative.
The term $\beta (c^a_0-c^a)$ represents actin turnover, with $\beta$ being the turnover rate and $c^a_0$ the equilibrium concentration in the cortex.
The term  $D^\mu\Delta^sc^\mu$ represents the surface diffusion of myosin, $D^\mu$ is the diffusion coefficient.
The average myosin concentration $c^\mu_0$ is conserved by eq. (\ref{advection2}).  

This model contains two different active drivings of cell motility: molecular motors are pulling agents generating positive stresses in \eqref{surfacestress} while  actin turnover in \eqref{advection1} can be associated with pushing agents generating negative stress in \eqref{surfacestress}. The interplay between these two types of agents in cell crawling was analyzed in \cite{Carlsson2011, Recho2015a}.

\paragraph{Results}
System (\ref{surfacestress})-(\ref{advection2}) constitutes a closed set of equations for determining the cortex velocity field and the actomyosin dynamics. 
%
The solution strategy is as follows: It can be shown (see \cite{SInfo}) that for a spherical cell the cortex flow is potential, $\boldsymbol u^c=\boldsymbol\nabla^s U+\boldsymbol v_s$, where $U$ is the flow potential.
The mechanical part of the problem \eqref{surfacestress}-\eqref{BIE} is then solved analytically by expansion  in scalar or vector spherical harmonics \cite{SInfo}.
This allows us to express $U$ for a spherical shape as a linear combination of the harmonic coefficients of $c^a$ and $c^\mu$ and compute the swimming velocity 
\begin{equation}
\label{vs}
v_s=\frac{2}{3}R\frac{\alpha c_1^a-\chi c_1^\mu}{2(\eta_s+\eta_b)+R(3\eta_{in}+2\eta_{out})},
\end{equation}
where $c_1^a$ and $c_1^\mu$ are the first harmonics of the actin and myosin concentration, defined precisely below.

Equations (\ref{advection1}) and (\ref{advection2}) are nonlinear but still can be tackled analytically by perturbation expansion.
We validated the analytical results by numerical solution of eqs. (\ref{advection1}) and (\ref{advection2}).
The details of the numerical method are given in \cite{SInfo}.

An important observation is that 
the dissipation mechanism highlighted in Eq.\ (\ref{vs}) arises from a combination of cortex, internal and
external fluid viscosities.  They  act as dashpots in parallel, as shown  by viscosity
additivity (up to numerical prefactors). Interestingly, the speed
is finite even if $\eta_{out}=0$: while the suspending fluid is
essential for swimming, its viscosity is not decisive for setting the value of
the swimming speed. 
We therefore consider the limit of $\eta_{in}=\eta_{out}=0$ in the rest of this
study, motivated by the fact that in physiological conditions most of the dissipation occurs in the cortex.
We also set $\eta_b=0$ for simplicity. Along with these assumptions,  we use from now on as characteristic scales $R$ for space, $R^2/D^\mu$ for time, $D^{\mu}\eta_s/R^2$ for surface stresses, $c_0^a$ and $c^{\mu}_a$ for actin and  myosin concentrations but keep the same symbols to avoid new notations. Three non-dimensional parameters fully define the problem,  the motor contractility $\bar\chi\equiv\chi c_0^\mu R^2/(D^\mu\eta_s)$, the compressibility of actin $\bar\alpha\equiv\alpha c_0^a R^2/(D^\mu\eta_s)$ and the turnover rate of actin $\bar\beta\equiv R^2\beta/D^\mu$. 



\begin{figure}
\begin{center}
\includegraphics[width=0.9\columnwidth]{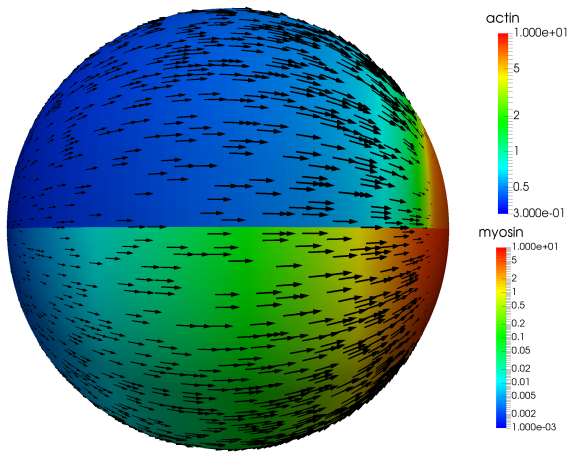}
\caption{\label{pattern} (Color on-line) The actin and myosin distribution (color fields) and cortex flow (arrows). Cell swims to the left.}
\end{center}
\end{figure}

\begin{figure}
\begin{center}
\includegraphics[width=0.9\columnwidth]{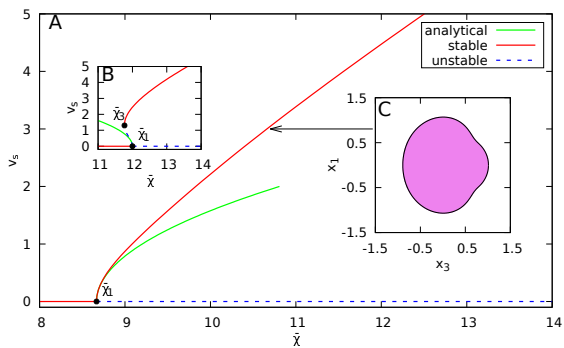}
\caption{\label{pitchfork} (Color on-line) A and B:  Swimming velocity as a function of contractility. Stable (unstable) branches are shown by continuous (dashed) lines. A: Supercritical case $\bar\alpha=10.0,$ $\bar\beta=3.0$. B: Subcritical case $\bar\alpha=10.0,$ $\bar\beta=2.0.$ Inset C shows a typical shape (retrograde flow is from left to right).}
\end{center}
\end{figure}

A homogeneous solution for actin and myosin ($c^a=c^{\mu}=1$) always exits.
This solution yields a zero cortical flow field and the cell is at rest. This solution can become unstable leading to a polarized cortex dynamics.
Such symmetry breaking results in a cortical flow and, in turn, in spontaneous cell motion.  

To quantify such process, we expand scalar fields in axisymmetric spherical harmonics as 
$$c^{a,\mu}(\theta)=1+\sum_{l=1}^\infty c^{a,\mu}_lP_l(\cos \theta),$$
 where $P_l$ are Legendre polynomials and $\theta$ is the polar angle between the swimming direction and 
the vector position on the sphere counted from the center of mass.
Linearizing eqs. (\ref{advection1}) and (\ref{advection2}) about the homogeneous solution, yields 
\begin{equation}\label{dyna_modes}
\dot{c}^a_l=J_l^{aa} c_l^a+J_l^{a\mu} c_l^\mu, \;\;\;  
\dot{c}^\mu_l=J_l^{\mu a} c_l^a+ J_l^{\mu\mu} c_l^\mu,
\end{equation}
where $J_l^{aa}=-l(l+1)\bar\alpha/\lambda_l-\bar\beta,$ $J_l^{a\mu}=l(l+1)\bar\chi/\lambda_l,$ $J_l^{\mu a}=-l(l+1)\bar\alpha/\lambda_l,$ $J_l^{\mu\mu}=l(l+1)\bar\chi/\lambda_l-l(l+1),$ and $\lambda_l=2(l^2+l-1)$.
The growth rate $\omega$ of a perturbation in \eqref{dyna_modes} therefore obeys a quadratic equation.

When the real part of $\omega$ vanishes (onset of instability), the imaginary part can either be zero or non zero.
The first case defines a steady bifurcation, while the second one defines a Hopf bifurcation.
A simple analysis shows that the minimum contractility at which the instability occurs  is controlled by the first harmonic $l=1$. 
The instability takes place if $\bar\chi>\bar\chi_c=\min(\bar\chi_1,\bar\chi_2),$ where
\begin{equation}
\label{critical}
\bar\chi_1=2+\frac{2\bar\alpha}{\bar\beta},\,\,\,\bar\chi_2=2+\bar\alpha+\bar\beta.
\end{equation}
We get a steady bifurcation for $\bar\chi_1<\bar\chi_2$ and a Hopf one otherwise. Once the  contractility exceeds the critical value given by (\ref{critical}) the perturbations grow exponentially with time and nonlinear terms can no longer be neglected. In general resorting to numerical resolution is necessary. However, in the vicinity of the bifurcation point a perturbation analysis is possible. 

First, let us consider the case of a steady bifurcation. A nonzero first harmonic  (concentration polarity) induces a spontaneous cell motion (See \eqref{vs}). This propulsion mode is driven by the appearance of a retrograde flow of actin, directed from a front region to a rear region and compensated by actin creation at the front ($\bar\beta (1-c^a)>0$, polymerization) and degradation at the rear ($\bar\beta (1-c^a)<0$, depolymerization). Figure \ref{pattern} shows the acto-myosin and the cortex flow patterns. The nature of this instability is  similar to the one presented in \cite{recho2013contraction} in a simplified picture.

The unstable eigenmode is given by  $C_1=2(\bar\alpha+\bar\beta) c_1^\mu-2\bar\alpha c_1^a$. 
To obtain the velocity close to the bifurcation threshold, we must also consider  the second harmonic.
We set $\dot{c}^a_2=\dot{c}^\mu_2=0$, because relaxation times for these modes are much smaller than that of $C_1$. This yields the expressions of $ c^a_2$ and $c^\mu_2$ as a function of $C_1,$ which upon  substitution  into the expression of $\dot C_1$ leads to 
\begin{equation}
\begin{aligned}
&\dot C_1=\frac{\bar\beta^2}{\bar\beta^2+\bar\alpha\bar\beta-2\bar\alpha}\left[(\bar\chi-\bar\chi_1)C_1+\nu C_1^3\right],\\
&\nu=-\frac{6\bar\alpha^2\bar\beta-12\bar\alpha^2+\bar\alpha\bar\beta^3+6\bar\alpha\bar\beta^2-24\bar\alpha\bar\beta+\bar\beta^4}{\bar\beta(\bar\alpha+2\bar\beta)(\bar\beta^2+\bar\alpha\bar\beta-2\bar\alpha)^2}.
\label{3order}
\end{aligned}
\end{equation}
The sign of $\nu$ dictates the nature of the steady bifurcation with $\nu<0$ corresponding to a supercritical  bifurcation (nonlinearity   saturates  linear growth) and $\nu>0$ corresponding to a subcritical one (nonlinearity amplifies  linear growth).

In the case of a supercritical bifurcation, \eqref{3order} has a stable fixed point given by $C_1=\sqrt{(\bar\chi-\bar\chi_1)/\nu}$. The swimming speed then reads
\begin{equation}
\label{vs2}
 v_s= \frac{\bar\beta}{3(\bar\beta^2+\bar\alpha\bar\beta-2\bar\alpha)}\sqrt{\frac{\bar\chi-\bar\chi_1}{-\nu}}.
\end{equation}
The bifurcation diagram in Fig. \ref{pitchfork}A shows both the analytical results and  the full numerical simulations. 

When the steady bifurcation is subcritical  (i.e. first order  transition),  velocity abruptly switches to a finite value. Due to the finite jump, the perturbative expansion leading to eq.(\ref{3order}) is illegitimate. One has thus to resort to a numerical solution. Figure~\ref{pitchfork}B shows the results. There is  a region of metastable coexistence between a non-polarized static state and a polarized motile one between a turning point located at $\bar\chi=\bar\chi_3$ and the linear stability threshold $\bar\chi=\bar\chi_1$. A finite perturbation at a constant contractility can therefore be sufficient to initiate or arrest motion in this range.

\begin{figure}
\begin{center}
\includegraphics[width=0.9\columnwidth]{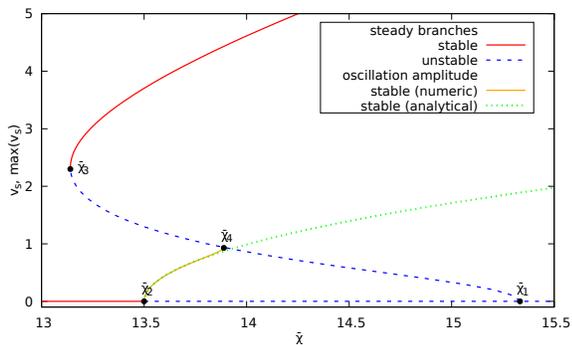}
\caption{\label{Hopf} (Color on-line) Amplitude of the swimming velocity as a function of contractility, supercritical Hopf case. $\bar\alpha=10.0,$ $\bar\beta=1.5$. Stable (unstable) branches are shown by continuous (dashed) lines. Dotted line shows maximum velocity for the limit cycle.
}
\end{center}
\end{figure}

When the bifurcation is of Hopf type, it is always supercritical with periodic oscillations of velocity emerging continuously from the static trivial solution at the critical value of contractility $\bar\chi_2$. A second critical contractility $\bar\chi_4$ exists at which the periodic oscillations disappear and the system abruptly jumps to a stable  motile steady-state.
The steady-state branch can be continued for $\bar\chi\ge\bar\chi_3.$
An unstable branch bent backwards emerges continuously from the uniform solution at critical contractility $\bar\chi_1,$ and annihilates with the stable steady-state branch.
The full bifurcation diagram shown in Fig.\ref{Hopf} is reminiscent of a heteroclinic bifurcation. In the range $[\bar\chi_3,\bar\chi_2]$, we again observe the coexistence of static and motile solutions while in the interval $[\bar\chi_3,\bar\chi_4]$, oscillatory and motile solutions are metastable potentially giving rise to complex cell gaits in the presence of biological noise. 

\begin{figure}
\begin{center}
\includegraphics[width=0.9\columnwidth]{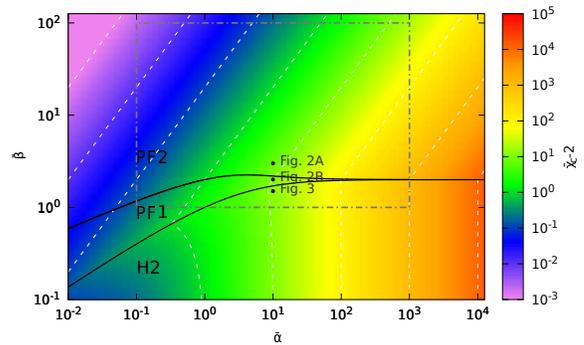}
\caption{\label{phd} (Color on-line) Phase diagram of bifurcation types as a function of $\bar\alpha$ and $\bar\beta.$ Color code shows (in log scale) the critical contractility $\bar\chi_c,$ for which the uniform solution loses stability. The bifurcation types are supercritical (PF2) and subcritical (PF1) pitchfork and supercritical Hopf (H2). Rectangle shows typical estimates of realistic parameters.}
\end{center}
\end{figure}

We summarize these results on the bifurcation nature in a phase diagram in Fig.~\ref{phd} where the critical value of $\bar\chi_c$ is shown in color code.


Relaxing the fixed shape assumption, we have determined the trend of the cell shape. For that purpose we have assumed the shape of the cell to be preserved by surface tension (see \cite{SInfo}). A typical shape is shown in Fig.\ref{pitchfork}C having a satisfactory tendency (mushroom-like flattened shape)  compared to some real cell shapes \cite{Ruprecht_2015_CCT}. 

\paragraph{Discussion}
A rich panel of patterns and dynamics due to a spontaneous symmetry breaking of the acto-myosin kinetics is revealed. The cell thus swims thanks to a retrograde flow of the cortex grasping on the external fluid. Orders of magnitudes of the three key parameters $\bar\alpha$, $\bar\beta$ and $\bar\chi$ can be estimated from available experimental data  $\bar\alpha\sim 10^{-1}-10^{3}$, $\bar\beta\sim 1-10^2$ and $\bar\chi\sim 10^{-2}-10^3$ (see \cite{SInfo}).  The wide ranges are due to disparate values of $\eta_s$ in the literature \cite{Wu+Wirtz.2018.1}.  The rectangle in Fig.~\ref{phd} shows the ranges of estimated parameters. It is interesting to see that this corresponds to a rich region of dynamics including the various bifurcations encountered in this study. The swimming speed (\ref{vs2}) yields (in the parameter range shown  by rectangle in Fig. \ref{phd})  in physical units about $v_s\sim 1-10 D^\mu/R\sim 0.1-1\;  \mu \text{m}/ \text{min}$, which is a reasonable value for many mammalian cells \cite{Jilkine2011}.

How the cortex flow is transmitted to the external fluid is still unclear. Three pathways are possible (i) through the membrane of the cell via recirculation of phospholipids, (ii) through transmembrane proteins (iii) through bumps that are advected backward by cortex in the form of waves (or shape changes). For axisymmetric flow (as considered here) the first scenario is impossible (it can be shown that the membrane is at rest in the swimming frame), except if endo/exocytosis is operating as reported recently \cite{jones2006endocytic}. The second scenario has been recently successfully applied to the T-lymphocyte swimming \cite{Theodoly}. The third scenario would require a cooperative motion of bumps along the membrane but numerical evaluation seems to rule out this possibility \cite{Theodoly}.

Our results show that shape-invariant actomyosin-based motility, a hallmark of
crawling, can occur away from any solid substrate, provided that the retrograde
flow of actin can transmit momentum to the surrounding fluid. This is consistent
with recent experimental observations \cite{Theodoly} and a theoretical model of active nematic droplets \cite{giomi2014spontaneous}. The rich bifurcation structure that allows
actomyosin to break symmetry is also reminiscent of the one known for cells crawling 
on a solid substrate \cite{recho2013contraction,Tjhung+Cates.2015.1, lavi2016deterministic}  pointing to the interpretation that the interaction of the actomyosin contractility with its
own viscous resistance to deformation and turn-over is sufficient to propel the cell robustly on a solid and in a fluid.

\begin{acknowledgments}\paragraph{Acknowledgments} We are grateful to Karin John for her role in the initial discussions leading to this work. AF and CM thank CNES (Centre National d'Etudes Spatiales) and the French-German University Programme ``Living Fluids'' (Grant CFDA-Q1-14) for financial support. JE and PR are supported by a CNRS Momentum grant, ANR-11-LABX-0030 ``Tec21'' and IRS ``AnisoTiss'' of Idex Univ.\ Grenoble Alpes.  All are members of GDR 3570 MecaBio and GDR 3070 CellTiss of CNRS. The computations were performed using the Cactus cluster of the CIMENT infrastructure, supported by the Rh\^one-Alpes region (GRANT CPER07\_13 CIRA) and the authors thank Philippe Beys who manages the cluster. 
\end{acknowledgments}
\bibliographystyle{apsrev}
\bibliography{crawlswim}

\begin{thebibliography}{46}
\expandafter\ifx\csname natexlab\endcsname\relax\def\natexlab#1{#1}\fi
\expandafter\ifx\csname bibnamefont\endcsname\relax
  \def\bibnamefont#1{#1}\fi
\expandafter\ifx\csname bibfnamefont\endcsname\relax
  \def\bibfnamefont#1{#1}\fi
\expandafter\ifx\csname citenamefont\endcsname\relax
  \def\citenamefont#1{#1}\fi
\expandafter\ifx\csname url\endcsname\relax
  \def\url#1{\texttt{#1}}\fi
\expandafter\ifx\csname urlprefix\endcsname\relax\def\urlprefix{URL }\fi
\providecommand{\bibinfo}[2]{#2}
\providecommand{\eprint}[2][]{\url{#2}}

\bibitem[{\citenamefont{Mogilner}(2009)}]{mogilner2009mathematics}
\bibinfo{author}{\bibfnamefont{A.}~\bibnamefont{Mogilner}},
  \bibinfo{journal}{Journal of mathematical biology}
  \textbf{\bibinfo{volume}{58}}, \bibinfo{pages}{105} (\bibinfo{year}{2009}).

\bibitem[{\citenamefont{Petrie and Yamada}(2015)}]{petrie2015fibroblasts}
\bibinfo{author}{\bibfnamefont{R.~J.} \bibnamefont{Petrie}} \bibnamefont{and}
  \bibinfo{author}{\bibfnamefont{K.~M.} \bibnamefont{Yamada}},
  \bibinfo{journal}{Trends in cell biology} \textbf{\bibinfo{volume}{25}},
  \bibinfo{pages}{666} (\bibinfo{year}{2015}).

\bibitem[{\citenamefont{Zhu and Mogilner}(2016)}]{zhu2016comparison}
\bibinfo{author}{\bibfnamefont{J.}~\bibnamefont{Zhu}} \bibnamefont{and}
  \bibinfo{author}{\bibfnamefont{A.}~\bibnamefont{Mogilner}},
  \bibinfo{journal}{Interface focus} \textbf{\bibinfo{volume}{6}},
  \bibinfo{pages}{20160040} (\bibinfo{year}{2016}).

\bibitem[{\citenamefont{Aroush et~al.}(2017)\citenamefont{Aroush, Ofer,
  Abu-Shah, Allard, Krichevsky, Mogilner, and Keren}}]{aroush2017actin}
\bibinfo{author}{\bibfnamefont{D.~R.-B.} \bibnamefont{Aroush}},
  \bibinfo{author}{\bibfnamefont{N.}~\bibnamefont{Ofer}},
  \bibinfo{author}{\bibfnamefont{E.}~\bibnamefont{Abu-Shah}},
  \bibinfo{author}{\bibfnamefont{J.}~\bibnamefont{Allard}},
  \bibinfo{author}{\bibfnamefont{O.}~\bibnamefont{Krichevsky}},
  \bibinfo{author}{\bibfnamefont{A.}~\bibnamefont{Mogilner}}, \bibnamefont{and}
  \bibinfo{author}{\bibfnamefont{K.}~\bibnamefont{Keren}},
  \bibinfo{journal}{Current Biology} \textbf{\bibinfo{volume}{27}},
  \bibinfo{pages}{2963} (\bibinfo{year}{2017}).

\bibitem[{\citenamefont{Paluch and Raz}(2013)}]{paluch2013role}
\bibinfo{author}{\bibfnamefont{E.~K.} \bibnamefont{Paluch}} \bibnamefont{and}
  \bibinfo{author}{\bibfnamefont{E.}~\bibnamefont{Raz}},
  \bibinfo{journal}{Current opinion in cell biology}
  \textbf{\bibinfo{volume}{25}}, \bibinfo{pages}{582} (\bibinfo{year}{2013}).

\bibitem[{\citenamefont{Lee et~al.}(1994)\citenamefont{Lee, Leonard, Oliver,
  Ishihara, and Jacobson}}]{Lee1957}
\bibinfo{author}{\bibfnamefont{J.}~\bibnamefont{Lee}},
  \bibinfo{author}{\bibfnamefont{M.}~\bibnamefont{Leonard}},
  \bibinfo{author}{\bibfnamefont{T.}~\bibnamefont{Oliver}},
  \bibinfo{author}{\bibfnamefont{A.}~\bibnamefont{Ishihara}}, \bibnamefont{and}
  \bibinfo{author}{\bibfnamefont{K.}~\bibnamefont{Jacobson}},
  \bibinfo{journal}{The Journal of Cell Biology}
  \textbf{\bibinfo{volume}{127}}, \bibinfo{pages}{1957} (\bibinfo{year}{1994}),
  ISSN \bibinfo{issn}{0021-9525}.

\bibitem[{\citenamefont{Rubinstein et~al.}(2009)\citenamefont{Rubinstein,
  Fournier, Jacobson, Verkhovsky, and
  Mogilner}}]{Rubinstein+Verkhovsky-Mogilner.2009.1}
\bibinfo{author}{\bibfnamefont{B.}~\bibnamefont{Rubinstein}},
  \bibinfo{author}{\bibfnamefont{M.~F.} \bibnamefont{Fournier}},
  \bibinfo{author}{\bibfnamefont{K.}~\bibnamefont{Jacobson}},
  \bibinfo{author}{\bibfnamefont{A.}~\bibnamefont{Verkhovsky}},
  \bibnamefont{and} \bibinfo{author}{\bibfnamefont{A.}~\bibnamefont{Mogilner}},
  \bibinfo{journal}{Biophys. J.} \textbf{\bibinfo{volume}{97}},
  \bibinfo{pages}{1853} (\bibinfo{year}{2009}).

\bibitem[{\citenamefont{Tjhung et~al.}(2015)\citenamefont{Tjhung, Tiribocchi,
  Marenduzzo, and Cates}}]{Tjhung+Cates.2015.1}
\bibinfo{author}{\bibfnamefont{E.}~\bibnamefont{Tjhung}},
  \bibinfo{author}{\bibfnamefont{A.}~\bibnamefont{Tiribocchi}},
  \bibinfo{author}{\bibfnamefont{D.}~\bibnamefont{Marenduzzo}},
  \bibnamefont{and} \bibinfo{author}{\bibfnamefont{M.~E.} \bibnamefont{Cates}},
  \bibinfo{journal}{Nat Commun} \textbf{\bibinfo{volume}{6}},
  \bibinfo{pages}{462} (\bibinfo{year}{2015}).

\bibitem[{\citenamefont{Ziebert et~al.}(2011)\citenamefont{Ziebert,
  Swaminathan, and Aranson}}]{ziebert2011model}
\bibinfo{author}{\bibfnamefont{F.}~\bibnamefont{Ziebert}},
  \bibinfo{author}{\bibfnamefont{S.}~\bibnamefont{Swaminathan}},
  \bibnamefont{and} \bibinfo{author}{\bibfnamefont{I.~S.}
  \bibnamefont{Aranson}}, \bibinfo{journal}{Journal of The Royal Society
  Interface} p. \bibinfo{pages}{rsif20110433} (\bibinfo{year}{2011}).

\bibitem[{\citenamefont{L{\"a}mmermann
  et~al.}(2008)\citenamefont{L{\"a}mmermann, Bader, Monkley, Worbs,
  Wedlich-S{\"o}ldner, Hirsch, Keller, F{\"o}rster, Critchley, F{\"a}ssler
  et~al.}}]{lammermann2008rapid}
\bibinfo{author}{\bibfnamefont{T.}~\bibnamefont{L{\"a}mmermann}},
  \bibinfo{author}{\bibfnamefont{B.~L.} \bibnamefont{Bader}},
  \bibinfo{author}{\bibfnamefont{S.~J.} \bibnamefont{Monkley}},
  \bibinfo{author}{\bibfnamefont{T.}~\bibnamefont{Worbs}},
  \bibinfo{author}{\bibfnamefont{R.}~\bibnamefont{Wedlich-S{\"o}ldner}},
  \bibinfo{author}{\bibfnamefont{K.}~\bibnamefont{Hirsch}},
  \bibinfo{author}{\bibfnamefont{M.}~\bibnamefont{Keller}},
  \bibinfo{author}{\bibfnamefont{R.}~\bibnamefont{F{\"o}rster}},
  \bibinfo{author}{\bibfnamefont{D.~R.} \bibnamefont{Critchley}},
  \bibinfo{author}{\bibfnamefont{R.}~\bibnamefont{F{\"a}ssler}},
  \bibnamefont{et~al.}, \bibinfo{journal}{Nature}
  \textbf{\bibinfo{volume}{453}}, \bibinfo{pages}{51} (\bibinfo{year}{2008}).

\bibitem[{\citenamefont{Bergert
  et~al.}(2015{\natexlab{a}})\citenamefont{Bergert, Anna, A., Aspalter, Oates,
  Charras, Salbreux, and Paluch}}]{Bergert2015}
\bibinfo{author}{\bibfnamefont{M.}~\bibnamefont{Bergert}},
  \bibinfo{author}{\bibfnamefont{E.}~\bibnamefont{Anna}},
  \bibinfo{author}{\bibfnamefont{D.~R.} \bibnamefont{A.}},
  \bibinfo{author}{\bibfnamefont{I.~M.} \bibnamefont{Aspalter}},
  \bibinfo{author}{\bibfnamefont{A.~C.} \bibnamefont{Oates}},
  \bibinfo{author}{\bibfnamefont{G.}~\bibnamefont{Charras}},
  \bibinfo{author}{\bibfnamefont{G.}~\bibnamefont{Salbreux}}, \bibnamefont{and}
  \bibinfo{author}{\bibfnamefont{E.~K.} \bibnamefont{Paluch}},
  \bibinfo{journal}{Nat. Cel. Biol.} \textbf{\bibinfo{volume}{17}},
  \bibinfo{pages}{524} (\bibinfo{year}{2015}{\natexlab{a}}).

\bibitem[{\citenamefont{Paluch et~al.}(2016)\citenamefont{Paluch, Aspalter, and
  Sixt}}]{paluch_review}
\bibinfo{author}{\bibfnamefont{E.~K.} \bibnamefont{Paluch}},
  \bibinfo{author}{\bibfnamefont{I.~M.} \bibnamefont{Aspalter}},
  \bibnamefont{and} \bibinfo{author}{\bibfnamefont{M.}~\bibnamefont{Sixt}},
  \bibinfo{journal}{Annual Review of Cell and Developmental Biology}
  \textbf{\bibinfo{volume}{32}}, \bibinfo{pages}{469} (\bibinfo{year}{2016}),
  \bibinfo{note}{pMID: 27501447}.

\bibitem[{\citenamefont{Renkawitz et~al.}(2009)\citenamefont{Renkawitz,
  Schumann, Weber, Laemmermann, Pflicke, Piel, Polleux, Spatz, and
  Sixt}}]{Renkawitz2009}
\bibinfo{author}{\bibfnamefont{J.}~\bibnamefont{Renkawitz}},
  \bibinfo{author}{\bibfnamefont{K.}~\bibnamefont{Schumann}},
  \bibinfo{author}{\bibfnamefont{M.}~\bibnamefont{Weber}},
  \bibinfo{author}{\bibfnamefont{T.}~\bibnamefont{Laemmermann}},
  \bibinfo{author}{\bibfnamefont{H.}~\bibnamefont{Pflicke}},
  \bibinfo{author}{\bibfnamefont{M.}~\bibnamefont{Piel}},
  \bibinfo{author}{\bibfnamefont{J.}~\bibnamefont{Polleux}},
  \bibinfo{author}{\bibfnamefont{J.~P.} \bibnamefont{Spatz}}, \bibnamefont{and}
  \bibinfo{author}{\bibfnamefont{M.}~\bibnamefont{Sixt}},
  \bibinfo{journal}{Nature Cell Biology} \textbf{\bibinfo{volume}{11}},
  \bibinfo{pages}{1438} (\bibinfo{year}{2009}).

\bibitem[{\citenamefont{Hawkins et~al.}(2009)\citenamefont{Hawkins, Piel,
  Faure-Andre, Lennon-Dumenil, Joanny, Prost, and Voituriez}}]{Hawkins2009}
\bibinfo{author}{\bibfnamefont{R.~J.} \bibnamefont{Hawkins}},
  \bibinfo{author}{\bibfnamefont{M.}~\bibnamefont{Piel}},
  \bibinfo{author}{\bibfnamefont{G.}~\bibnamefont{Faure-Andre}},
  \bibinfo{author}{\bibfnamefont{A.~M.} \bibnamefont{Lennon-Dumenil}},
  \bibinfo{author}{\bibfnamefont{J.~F.} \bibnamefont{Joanny}},
  \bibinfo{author}{\bibfnamefont{J.}~\bibnamefont{Prost}}, \bibnamefont{and}
  \bibinfo{author}{\bibfnamefont{R.}~\bibnamefont{Voituriez}},
  \bibinfo{journal}{Phys. Rev. Lett.} \textbf{\bibinfo{volume}{102}},
  \bibinfo{pages}{058103} (\bibinfo{year}{2009}).

\bibitem[{\citenamefont{Barry and Bretscher}(2010)}]{barry2010dictyostelium}
\bibinfo{author}{\bibfnamefont{N.~P.} \bibnamefont{Barry}} \bibnamefont{and}
  \bibinfo{author}{\bibfnamefont{M.~S.} \bibnamefont{Bretscher}},
  \bibinfo{journal}{Proceedings of the National Academy of Sciences}
  \textbf{\bibinfo{volume}{107}}, \bibinfo{pages}{11376}
  (\bibinfo{year}{2010}).

\bibitem[{\citenamefont{Arroyo et~al.}(2012)\citenamefont{Arroyo, Heltai,
  Mill{\'a}n, and DeSimone}}]{arroyo2012reverse}
\bibinfo{author}{\bibfnamefont{M.}~\bibnamefont{Arroyo}},
  \bibinfo{author}{\bibfnamefont{L.}~\bibnamefont{Heltai}},
  \bibinfo{author}{\bibfnamefont{D.}~\bibnamefont{Mill{\'a}n}},
  \bibnamefont{and} \bibinfo{author}{\bibfnamefont{A.}~\bibnamefont{DeSimone}},
  \bibinfo{journal}{Proceedings of the National Academy of Sciences}
  \textbf{\bibinfo{volume}{109}}, \bibinfo{pages}{17874}
  (\bibinfo{year}{2012}).

\bibitem[{\citenamefont{Aoun et~al.}(2019)\citenamefont{Aoun, Negre, Farutin,
  Garcia-Seyda, Rivzi, Galland, Michelot, Luo, Biarnes-Pelicot, Hivroz
  et~al.}}]{Theodoly}
\bibinfo{author}{\bibfnamefont{L.}~\bibnamefont{Aoun}},
  \bibinfo{author}{\bibfnamefont{P.}~\bibnamefont{Negre}},
  \bibinfo{author}{\bibfnamefont{A.}~\bibnamefont{Farutin}},
  \bibinfo{author}{\bibfnamefont{N.}~\bibnamefont{Garcia-Seyda}},
  \bibinfo{author}{\bibfnamefont{M.~S.} \bibnamefont{Rivzi}},
  \bibinfo{author}{\bibfnamefont{R.}~\bibnamefont{Galland}},
  \bibinfo{author}{\bibfnamefont{A.}~\bibnamefont{Michelot}},
  \bibinfo{author}{\bibfnamefont{X.}~\bibnamefont{Luo}},
  \bibinfo{author}{\bibfnamefont{M.}~\bibnamefont{Biarnes-Pelicot}},
  \bibinfo{author}{\bibfnamefont{C.}~\bibnamefont{Hivroz}},
  \bibnamefont{et~al.}, \bibinfo{journal}{bioRxiv}  (\bibinfo{year}{2019}).

\bibitem[{\citenamefont{Lauga and Powers}(2009)}]{lauga2009hydrodynamics}
\bibinfo{author}{\bibfnamefont{E.}~\bibnamefont{Lauga}} \bibnamefont{and}
  \bibinfo{author}{\bibfnamefont{T.~R.} \bibnamefont{Powers}},
  \bibinfo{journal}{Reports on Progress in Physics}
  \textbf{\bibinfo{volume}{72}}, \bibinfo{pages}{096601}
  (\bibinfo{year}{2009}).

\bibitem[{\citenamefont{Stone and Samuel}(1996)}]{stone1996propulsion}
\bibinfo{author}{\bibfnamefont{H.~A.} \bibnamefont{Stone}} \bibnamefont{and}
  \bibinfo{author}{\bibfnamefont{A.~D.~T.} \bibnamefont{Samuel}},
  \bibinfo{journal}{Physical review letters} \textbf{\bibinfo{volume}{77}},
  \bibinfo{pages}{4102} (\bibinfo{year}{1996}).

\bibitem[{\citenamefont{Brennen and Winet}(1977)}]{brennen1977fluid}
\bibinfo{author}{\bibfnamefont{C.}~\bibnamefont{Brennen}} \bibnamefont{and}
  \bibinfo{author}{\bibfnamefont{H.}~\bibnamefont{Winet}},
  \bibinfo{journal}{Annual Review of Fluid Mechanics}
  \textbf{\bibinfo{volume}{9}}, \bibinfo{pages}{339} (\bibinfo{year}{1977}).

\bibitem[{\citenamefont{Fackler and Grosse}(2008)}]{fackler2008cell}
\bibinfo{author}{\bibfnamefont{O.~T.} \bibnamefont{Fackler}} \bibnamefont{and}
  \bibinfo{author}{\bibfnamefont{R.}~\bibnamefont{Grosse}},
  \bibinfo{journal}{The Journal of cell biology}
  \textbf{\bibinfo{volume}{181}}, \bibinfo{pages}{879} (\bibinfo{year}{2008}).

\bibitem[{\citenamefont{Farutin et~al.}(2013)\citenamefont{Farutin, Rafa\"{\i},
  Dysthe, Duperray, Peyla, and Misbah}}]{Farutin2013}
\bibinfo{author}{\bibfnamefont{A.}~\bibnamefont{Farutin}},
  \bibinfo{author}{\bibfnamefont{S.}~\bibnamefont{Rafa\"{\i}}},
  \bibinfo{author}{\bibfnamefont{D.~K.} \bibnamefont{Dysthe}},
  \bibinfo{author}{\bibfnamefont{A.}~\bibnamefont{Duperray}},
  \bibinfo{author}{\bibfnamefont{P.}~\bibnamefont{Peyla}}, \bibnamefont{and}
  \bibinfo{author}{\bibfnamefont{C.}~\bibnamefont{Misbah}},
  \bibinfo{journal}{Phys. Rev. Lett.} \textbf{\bibinfo{volume}{111}},
  \bibinfo{pages}{228102} (\bibinfo{year}{2013}).

\bibitem[{\citenamefont{Campbell and Bagchi}(2017)}]{campbell2017computational}
\bibinfo{author}{\bibfnamefont{E.~J.} \bibnamefont{Campbell}} \bibnamefont{and}
  \bibinfo{author}{\bibfnamefont{P.}~\bibnamefont{Bagchi}},
  \bibinfo{journal}{Physics of Fluids} \textbf{\bibinfo{volume}{29}},
  \bibinfo{pages}{101902} (\bibinfo{year}{2017}).

\bibitem[{\citenamefont{Verkhovsky et~al.}(1999)\citenamefont{Verkhovsky,
  Svitkina, and Borisy}}]{verkhovsky1999self}
\bibinfo{author}{\bibfnamefont{A.~B.} \bibnamefont{Verkhovsky}},
  \bibinfo{author}{\bibfnamefont{T.~M.} \bibnamefont{Svitkina}},
  \bibnamefont{and} \bibinfo{author}{\bibfnamefont{G.~G.}
  \bibnamefont{Borisy}}, \bibinfo{journal}{Current Biology}
  \textbf{\bibinfo{volume}{9}}, \bibinfo{pages}{11} (\bibinfo{year}{1999}).

\bibitem[{\citenamefont{Recho et~al.}(2013)\citenamefont{Recho, Putelat, and
  Truskinovsky}}]{recho2013contraction}
\bibinfo{author}{\bibfnamefont{P.}~\bibnamefont{Recho}},
  \bibinfo{author}{\bibfnamefont{T.}~\bibnamefont{Putelat}}, \bibnamefont{and}
  \bibinfo{author}{\bibfnamefont{L.}~\bibnamefont{Truskinovsky}},
  \bibinfo{journal}{Physical review letters} \textbf{\bibinfo{volume}{111}},
  \bibinfo{pages}{108102} (\bibinfo{year}{2013}).

\bibitem[{\citenamefont{Barnhart et~al.}(2015)\citenamefont{Barnhart, Lee,
  Allen, Theriot, and Mogilner}}]{barnhart2015balance}
\bibinfo{author}{\bibfnamefont{E.}~\bibnamefont{Barnhart}},
  \bibinfo{author}{\bibfnamefont{K.-C.} \bibnamefont{Lee}},
  \bibinfo{author}{\bibfnamefont{G.~M.} \bibnamefont{Allen}},
  \bibinfo{author}{\bibfnamefont{J.~A.} \bibnamefont{Theriot}},
  \bibnamefont{and} \bibinfo{author}{\bibfnamefont{A.}~\bibnamefont{Mogilner}},
  \bibinfo{journal}{Proceedings of the National Academy of Sciences} p.
  \bibinfo{pages}{201417257} (\bibinfo{year}{2015}).

\bibitem[{\citenamefont{Callan-Jones et~al.}(2016)\citenamefont{Callan-Jones,
  Ruprecht, Wieser, Heisenberg, and Voituriez}}]{Voituriez2016}
\bibinfo{author}{\bibfnamefont{A.~C.} \bibnamefont{Callan-Jones}},
  \bibinfo{author}{\bibfnamefont{V.}~\bibnamefont{Ruprecht}},
  \bibinfo{author}{\bibfnamefont{S.}~\bibnamefont{Wieser}},
  \bibinfo{author}{\bibfnamefont{C.~P.} \bibnamefont{Heisenberg}},
  \bibnamefont{and}
  \bibinfo{author}{\bibfnamefont{R.}~\bibnamefont{Voituriez}},
  \bibinfo{journal}{Phys. Rev. Lett.} \textbf{\bibinfo{volume}{116}},
  \bibinfo{pages}{028102} (\bibinfo{year}{2016}).

\bibitem[{\citenamefont{Lavi et~al.}(2016)\citenamefont{Lavi, Piel,
  Lennon-Dum{\'e}nil, Voituriez, and Gov}}]{lavi2016deterministic}
\bibinfo{author}{\bibfnamefont{I.}~\bibnamefont{Lavi}},
  \bibinfo{author}{\bibfnamefont{M.}~\bibnamefont{Piel}},
  \bibinfo{author}{\bibfnamefont{A.-M.} \bibnamefont{Lennon-Dum{\'e}nil}},
  \bibinfo{author}{\bibfnamefont{R.}~\bibnamefont{Voituriez}},
  \bibnamefont{and} \bibinfo{author}{\bibfnamefont{N.~S.} \bibnamefont{Gov}},
  \bibinfo{journal}{Nature Physics} \textbf{\bibinfo{volume}{12}},
  \bibinfo{pages}{1146} (\bibinfo{year}{2016}).

\bibitem[{\citenamefont{Godeau}(2016)}]{godeau2016cyclic}
\bibinfo{author}{\bibfnamefont{A.}~\bibnamefont{Godeau}}, Ph.D. thesis,
  \bibinfo{school}{Strasbourg} (\bibinfo{year}{2016}).

\bibitem[{\citenamefont{Kruse et~al.}(2004)\citenamefont{Kruse, Joanny,
  J{\"u}licher, Prost, and Sekimoto}}]{Kruse+.2004.1}
\bibinfo{author}{\bibfnamefont{K.}~\bibnamefont{Kruse}},
  \bibinfo{author}{\bibfnamefont{J.-F.} \bibnamefont{Joanny}},
  \bibinfo{author}{\bibfnamefont{F.}~\bibnamefont{J{\"u}licher}},
  \bibinfo{author}{\bibfnamefont{J.}~\bibnamefont{Prost}}, \bibnamefont{and}
  \bibinfo{author}{\bibfnamefont{K.}~\bibnamefont{Sekimoto}},
  \bibinfo{journal}{Phys. Rev. Lett.} \textbf{\bibinfo{volume}{92}},
  \bibinfo{pages}{078101} (\bibinfo{year}{2004}).

\bibitem[{\citenamefont{Farutin et~al.}(2012)\citenamefont{Farutin, Aouane, and
  Misbah}}]{Farutin2012}
\bibinfo{author}{\bibfnamefont{A.}~\bibnamefont{Farutin}},
  \bibinfo{author}{\bibfnamefont{O.}~\bibnamefont{Aouane}}, \bibnamefont{and}
  \bibinfo{author}{\bibfnamefont{C.}~\bibnamefont{Misbah}},
  \bibinfo{journal}{Phys. Rev. E} \textbf{\bibinfo{volume}{85}},
  \bibinfo{pages}{061922} (\bibinfo{year}{2012}).

\bibitem[{\citenamefont{Pozrikidis}(1992)}]{Pozrikidis1992}
\bibinfo{author}{\bibfnamefont{C.}~\bibnamefont{Pozrikidis}},
  \emph{\bibinfo{title}{Boundary Integral and Singularity Methods for
  Linearized Viscous Flow}} (\bibinfo{publisher}{Cambridge University Press,
  Cambridge, UK}, \bibinfo{year}{1992}).

\bibitem[{\citenamefont{AF et~al.}(2019)\citenamefont{AF, JE, CM, and
  PR}}]{SInfo}
\bibinfo{author}{\bibnamefont{AF}}, \bibinfo{author}{\bibnamefont{JE}},
  \bibinfo{author}{\bibnamefont{CM}}, \bibnamefont{and}
  \bibinfo{author}{\bibnamefont{PR}}, \bibinfo{journal}{Supplementary
  Information}  (\bibinfo{year}{2019}).

\bibitem[{\citenamefont{Carlsson}({2011})}]{Carlsson2011}
\bibinfo{author}{\bibfnamefont{A.~E.} \bibnamefont{Carlsson}},
  \bibinfo{journal}{{New Journal of Physics}} \textbf{\bibinfo{volume}{{13}}}
  (\bibinfo{year}{{2011}}), ISSN \bibinfo{issn}{{1367-2630}}.

\bibitem[{\citenamefont{Recho and Truskinovsky}(2015)}]{Recho2015a}
\bibinfo{author}{\bibfnamefont{P.}~\bibnamefont{Recho}} \bibnamefont{and}
  \bibinfo{author}{\bibfnamefont{L.}~\bibnamefont{Truskinovsky}},
  \bibinfo{journal}{Mathematics and Mechanics of Solids}
  (\bibinfo{year}{2015}).

\bibitem[{\citenamefont{Ruprecht et~al.}(2015)\citenamefont{Ruprecht, Wieser,
  Callan-Jones, Smutny, Morita, Sako, Barone, Ritsch-Marte, Sixt, Voituriez
  et~al.}}]{Ruprecht_2015_CCT}
\bibinfo{author}{\bibfnamefont{V.}~\bibnamefont{Ruprecht}},
  \bibinfo{author}{\bibfnamefont{S.}~\bibnamefont{Wieser}},
  \bibinfo{author}{\bibfnamefont{A.}~\bibnamefont{Callan-Jones}},
  \bibinfo{author}{\bibfnamefont{M.}~\bibnamefont{Smutny}},
  \bibinfo{author}{\bibfnamefont{H.}~\bibnamefont{Morita}},
  \bibinfo{author}{\bibfnamefont{K.}~\bibnamefont{Sako}},
  \bibinfo{author}{\bibfnamefont{V.}~\bibnamefont{Barone}},
  \bibinfo{author}{\bibfnamefont{M.}~\bibnamefont{Ritsch-Marte}},
  \bibinfo{author}{\bibfnamefont{M.}~\bibnamefont{Sixt}},
  \bibinfo{author}{\bibfnamefont{R.}~\bibnamefont{Voituriez}},
  \bibnamefont{et~al.}, \bibinfo{journal}{CELL} \textbf{\bibinfo{volume}{160}}
  (\bibinfo{year}{2015}).

\bibitem[{\citenamefont{Wu et~al.}(2018)\citenamefont{Wu, Aroush, Asnacios,
  Chen, Dokukin, Doss, Durand-Smet, Ekpenyong, Guck, Guz
  et~al.}}]{Wu+Wirtz.2018.1}
\bibinfo{author}{\bibfnamefont{P.-H.} \bibnamefont{Wu}},
  \bibinfo{author}{\bibfnamefont{D.~R.-B.} \bibnamefont{Aroush}},
  \bibinfo{author}{\bibfnamefont{A.}~\bibnamefont{Asnacios}},
  \bibinfo{author}{\bibfnamefont{W.-C.} \bibnamefont{Chen}},
  \bibinfo{author}{\bibfnamefont{M.~E.} \bibnamefont{Dokukin}},
  \bibinfo{author}{\bibfnamefont{B.~L.} \bibnamefont{Doss}},
  \bibinfo{author}{\bibfnamefont{P.}~\bibnamefont{Durand-Smet}},
  \bibinfo{author}{\bibfnamefont{A.}~\bibnamefont{Ekpenyong}},
  \bibinfo{author}{\bibfnamefont{J.}~\bibnamefont{Guck}},
  \bibinfo{author}{\bibfnamefont{N.~V.} \bibnamefont{Guz}},
  \bibnamefont{et~al.}, \bibinfo{journal}{Nat Methods}
  \textbf{\bibinfo{volume}{15}}, \bibinfo{pages}{491} (\bibinfo{year}{2018}).

\bibitem[{\citenamefont{Jilkine and Edelstein-Keshet}(2011)}]{Jilkine2011}
\bibinfo{author}{\bibfnamefont{A.}~\bibnamefont{Jilkine}} \bibnamefont{and}
  \bibinfo{author}{\bibfnamefont{L.}~\bibnamefont{Edelstein-Keshet}},
  \bibinfo{journal}{PLoS Comput Biol} \textbf{\bibinfo{volume}{7}},
  \bibinfo{pages}{e1001121} (\bibinfo{year}{2011}).

\bibitem[{\citenamefont{Jones et~al.}(2006)\citenamefont{Jones, Caswell, and
  Norman}}]{jones2006endocytic}
\bibinfo{author}{\bibfnamefont{M.~C.} \bibnamefont{Jones}},
  \bibinfo{author}{\bibfnamefont{P.~T.} \bibnamefont{Caswell}},
  \bibnamefont{and} \bibinfo{author}{\bibfnamefont{J.~C.}
  \bibnamefont{Norman}}, \bibinfo{journal}{Current opinion in cell biology}
  \textbf{\bibinfo{volume}{18}}, \bibinfo{pages}{549} (\bibinfo{year}{2006}).

\bibitem[{\citenamefont{Giomi and DeSimone}(2014)}]{giomi2014spontaneous}
\bibinfo{author}{\bibfnamefont{L.}~\bibnamefont{Giomi}} \bibnamefont{and}
  \bibinfo{author}{\bibfnamefont{A.}~\bibnamefont{DeSimone}},
  \bibinfo{journal}{Physical review letters} \textbf{\bibinfo{volume}{112}},
  \bibinfo{pages}{147802} (\bibinfo{year}{2014}).

\bibitem[{\citenamefont{Clark et~al.}(2013)\citenamefont{Clark, Dierkes, and
  Paluch}}]{clark2013monitoring}
\bibinfo{author}{\bibfnamefont{A.~G.} \bibnamefont{Clark}},
  \bibinfo{author}{\bibfnamefont{K.}~\bibnamefont{Dierkes}}, \bibnamefont{and}
  \bibinfo{author}{\bibfnamefont{E.~K.} \bibnamefont{Paluch}},
  \bibinfo{journal}{Biophysical journal} \textbf{\bibinfo{volume}{105}},
  \bibinfo{pages}{570} (\bibinfo{year}{2013}).

\bibitem[{\citenamefont{Turlier et~al.}(2014)\citenamefont{Turlier, Audoly,
  Prost, and Joanny}}]{turlier2014furrow}
\bibinfo{author}{\bibfnamefont{H.}~\bibnamefont{Turlier}},
  \bibinfo{author}{\bibfnamefont{B.}~\bibnamefont{Audoly}},
  \bibinfo{author}{\bibfnamefont{J.}~\bibnamefont{Prost}}, \bibnamefont{and}
  \bibinfo{author}{\bibfnamefont{J.-F.} \bibnamefont{Joanny}},
  \bibinfo{journal}{Biophysical journal} \textbf{\bibinfo{volume}{106}},
  \bibinfo{pages}{114} (\bibinfo{year}{2014}).

\bibitem[{\citenamefont{Bergert
  et~al.}(2015{\natexlab{b}})\citenamefont{Bergert, Erzberger, Desai, Aspalter,
  Oates, Charras, Salbreux, and Paluch}}]{bergert2015force}
\bibinfo{author}{\bibfnamefont{M.}~\bibnamefont{Bergert}},
  \bibinfo{author}{\bibfnamefont{A.}~\bibnamefont{Erzberger}},
  \bibinfo{author}{\bibfnamefont{R.~A.} \bibnamefont{Desai}},
  \bibinfo{author}{\bibfnamefont{I.~M.} \bibnamefont{Aspalter}},
  \bibinfo{author}{\bibfnamefont{A.~C.} \bibnamefont{Oates}},
  \bibinfo{author}{\bibfnamefont{G.}~\bibnamefont{Charras}},
  \bibinfo{author}{\bibfnamefont{G.}~\bibnamefont{Salbreux}}, \bibnamefont{and}
  \bibinfo{author}{\bibfnamefont{E.~K.} \bibnamefont{Paluch}},
  \bibinfo{journal}{Nature cell biology} \textbf{\bibinfo{volume}{17}},
  \bibinfo{pages}{524} (\bibinfo{year}{2015}{\natexlab{b}}).

\bibitem[{\citenamefont{Hawkins et~al.}(2011)\citenamefont{Hawkins, Poincloux,
  B{\'e}nichou, Piel, Chavrier, and Voituriez}}]{hawkins2011spontaneous}
\bibinfo{author}{\bibfnamefont{R.~J.} \bibnamefont{Hawkins}},
  \bibinfo{author}{\bibfnamefont{R.}~\bibnamefont{Poincloux}},
  \bibinfo{author}{\bibfnamefont{O.}~\bibnamefont{B{\'e}nichou}},
  \bibinfo{author}{\bibfnamefont{M.}~\bibnamefont{Piel}},
  \bibinfo{author}{\bibfnamefont{P.}~\bibnamefont{Chavrier}}, \bibnamefont{and}
  \bibinfo{author}{\bibfnamefont{R.}~\bibnamefont{Voituriez}},
  \bibinfo{journal}{Biophysical journal} \textbf{\bibinfo{volume}{101}},
  \bibinfo{pages}{1041} (\bibinfo{year}{2011}).

\bibitem[{\citenamefont{Uehara et~al.}(2010)\citenamefont{Uehara, Goshima,
  Mabuchi, Vale, Spudich, and Griffis}}]{uehara2010determinants}
\bibinfo{author}{\bibfnamefont{R.}~\bibnamefont{Uehara}},
  \bibinfo{author}{\bibfnamefont{G.}~\bibnamefont{Goshima}},
  \bibinfo{author}{\bibfnamefont{I.}~\bibnamefont{Mabuchi}},
  \bibinfo{author}{\bibfnamefont{R.~D.} \bibnamefont{Vale}},
  \bibinfo{author}{\bibfnamefont{J.~A.} \bibnamefont{Spudich}},
  \bibnamefont{and} \bibinfo{author}{\bibfnamefont{E.~R.}
  \bibnamefont{Griffis}}, \bibinfo{journal}{Current biology}
  \textbf{\bibinfo{volume}{20}}, \bibinfo{pages}{1080} (\bibinfo{year}{2010}).

\bibitem[{\citenamefont{Fritzsche et~al.}(2013)\citenamefont{Fritzsche,
  Lewalle, Duke, Kruse, and Charras}}]{fritzsche2013analysis}
\bibinfo{author}{\bibfnamefont{M.}~\bibnamefont{Fritzsche}},
  \bibinfo{author}{\bibfnamefont{A.}~\bibnamefont{Lewalle}},
  \bibinfo{author}{\bibfnamefont{T.}~\bibnamefont{Duke}},
  \bibinfo{author}{\bibfnamefont{K.}~\bibnamefont{Kruse}}, \bibnamefont{and}
  \bibinfo{author}{\bibfnamefont{G.}~\bibnamefont{Charras}},
  \bibinfo{journal}{Molecular biology of the cell}
  \textbf{\bibinfo{volume}{24}}, \bibinfo{pages}{757} (\bibinfo{year}{2013}).

\end{thebibliography}

\onecolumngrid

\section*{Supplementary information}

The supplementary information contains the explicit expressions for the Green's kernels of the Stokes equations, the calculation of the cortex flow in terms of the concentration fields, the description of the numerical procedure, the technique of shape reconstruction in the quasi-spherical limit and the physical values used to estimate non-dimensional parameters.

\section{Green's kernels}

The following kernels are to be used in eq. (2) of the main text:

\begin{align}
\label{kernels}
G_{ij}(\boldsymbol x,\boldsymbol x')&=\frac{1}{8\pi}\left[\frac{\delta_{ij}}{|\boldsymbol x-\boldsymbol x')|}+\frac{(\boldsymbol x-\boldsymbol x')_i(\boldsymbol x-\boldsymbol x')_j}{|\boldsymbol x-\boldsymbol x'|^3}\right],\notag\\
K_{ijk}(\boldsymbol x,\boldsymbol x')&=\frac{3}{4\pi}\frac{(\boldsymbol x-\boldsymbol x')_i(\boldsymbol x-\boldsymbol x')_j(\boldsymbol x-\boldsymbol x')_k}{|\boldsymbol x-\boldsymbol x'|^5}.
\end{align}

\section{Full solution}
The results in this section are presented in the most general form, without any assumptions about the relative values of $\eta_{in}$, $\eta_{out}$, $\eta_s$, and $\eta_b$.

\subsection{Spherical Harmonics}
All fields on the cell surface are expanded in spherical harmonics of the vector pointing from the center of the cell to a given point of its surface.
The spherical harmonics are defined as
\begin{equation}
\label{Y}
Y_{l,m}(\boldsymbol x)=\sqrt{\frac{2l+1}{4\pi}\frac{(l-m)!}{(l+m)!}}P_l^m\left(\frac{x_3}{x}\right)\left(\frac{x_1+ix_2}{|x_1+ix_2|}\right)^m,
\end{equation}
where $P_l^m$ are associated Legendre polynomials.
The following expansions are used
\begin{equation}
\label{sphericalca}
\begin{aligned}
&c^a(\boldsymbol x)=\sum\limits_{l=0}^\infty c^a_l(\boldsymbol x)\\
&c^a_l(\boldsymbol x)=\sum\limits_{m=-l}^lc^a_{l,m}Y_{l,m}(\boldsymbol x)\\
\end{aligned}
\end{equation}
\begin{equation}
\label{sphericalcm}
\begin{aligned}
&c^\mu(\boldsymbol x)=\sum\limits_{l=0}^\infty c^\mu_l(\boldsymbol x)\\
&c^\mu_l(\boldsymbol x)=\sum\limits_{m=-l}^lc^\mu_{l,m}Y_{l,m}(\boldsymbol x)\\
\end{aligned}
\end{equation}
\begin{equation}
\label{sphericalfn}
\begin{aligned}
&f^n(\boldsymbol x)=\sum\limits_{l=2}^\infty f^n_l(\boldsymbol x)\\
&f^n_l(\boldsymbol x)=\sum\limits_{m=-l}^lf^n_{l,m}Y_{l,m}(\boldsymbol x)\\
\end{aligned}
\end{equation}
\begin{equation}
\label{sphericalU}
\begin{aligned}
&U(\boldsymbol x)=\sum\limits_{l=1}^\infty U_l(\boldsymbol x)\\
&U_l(\boldsymbol x)=\sum\limits_{m=-l}^lU_{l,m}Y_{l,m}(\boldsymbol x)\\
\end{aligned}
\end{equation}

Note that $f^n_1=0$ is zero, as required by the condition of the total force acting on the cell being equal to zero, $f^n_0$ is irrelevant because it just shifts the osmotic pressure drop across the membrane, and $U_0$ is irrelevant because only gradients of $U$ enter equations.

We define the vector spherical harmonics as
\begin{equation}
\label{Y1}
\boldsymbol Y_{1,l,m} (\boldsymbol x) = [\boldsymbol \nabla^s -(l+1)\boldsymbol x] Y_{l,m}(\boldsymbol x)
\end{equation}
\begin{equation}
\label{Y2}
\boldsymbol Y_{2,l,m} (\boldsymbol x) = [\boldsymbol \nabla^s +l\boldsymbol x] Y_{l,m}(\boldsymbol x)
\end{equation}
\begin{equation}
\label{Y3}
\boldsymbol Y_{3,l,m} (\boldsymbol x) = \boldsymbol x \times \boldsymbol\nabla^s Y_{l,m}(\boldsymbol x)
\end{equation}
The following expansions are used
\begin{equation}
\label{vY}
\boldsymbol u^c=\sum\limits_{j=1}^3\sum\limits_{l=0}^\infty\sum\limits_{m=-l}^lu^c_{j,l,m}\boldsymbol Y_{j,l,m}(\boldsymbol x)
\end{equation}
\begin{equation}
\label{fY}
\boldsymbol f=\sum\limits_{j=1}^3\sum\limits_{l=0}^\infty\sum\limits_{m=-l}^lf_{j,l,m}\boldsymbol Y_{j,l,m}(\boldsymbol x)
\end{equation}

\subsection{Force calculation}
The force $\boldsymbol f$ can be represented as a sum of two contributions $f_{j,l,m}=f^v_{j,l,m}+f^e_{j,l,m},$ where
\begin{equation}
\label{f}
\begin{aligned}
&f^v_{1,l,m}=-\frac{(l+2)[2\eta_s(l^2+l+1)+\eta_b(l+1)(l+2)]}{2l+1}\frac{u^c_{1,l,m}}{R^2}-\frac{l(l-1)(l+2)(2\eta_s+\eta_b)}{2l+1}\frac{u^c_{2,l,m}}{R^2},\\
&f^v_{2,l,m}=-\frac{(l+1)(l+2)(l-1)(2\eta_s+\eta_b)}{2l+1}\frac{u^c_{1,l,m}}{R^2}-\frac{(l-1)[2\eta_s(l^2+l+1)+\eta_b(l-1)l]}{2l+1}\frac{u^c_{2,l,m}}{R^2},\\
&f^v_{3,l,m}=-\eta_s(l+2)(l-1)\frac{u^c_{3,l,m}}{R^2},\\
&f^e_{1,l,m}=\frac{l+2}{2l+1}\frac{\left[\chi c^\mu-\alpha c^a\right]_{l,m}}{R}+\frac{f^n_{l,m}}{2l+1},\\
&f^e_{2,l,m}=\frac{l-1}{2l+1}\frac{\left[\chi c^\mu-\alpha c^a\right]_{l,m}}{R}-\frac{f^n_{l,m}}{2l+1},\\
&f^e_{3,l,m}=0.
\end{aligned}
\end{equation}
The amplitudes $f^v_{j,l,m}$ contain the contribution of the surface viscosity terms in eq. (1) of the main text, while the amplitudes $f^e_{j,l,m}$ contain contributions of myosin contractility, cortex compressibility, and Lagrange multiplier $f^n\boldsymbol n$.

\subsection{Fluid dynamics}
The integrals in eq. (2) of the main text can be calculated analytically for a spherical cell.
The results are
\begin{equation}
\label{BIEsphere}
\frac{\eta_{in}+\eta_{out}}{2}u^c_{j,l,m}=Rg_{j,l}f_{j,l,m}+(\eta_{out}-\eta_{in})k_{j,l}u^c_{j,l,m},
\end{equation}
where the coefficients $g_{j,l}$ and $k_{j,l}$ are listed in table \ref{greentable}.

\begin{table}
\begin{center}
\begin{tabular}{c|ccc}
$j$ & 1 & 2 & 3 \\
\hline
$g_{j,l}$ & $\frac{l}{(2l+1)(2l+3)}$ & $\frac{l+1}{(2l-1)(2l+1)}$ & $\frac{1}{2l+1}$ \\
$k_{j,l}$ & $\frac{3}{2(2l+1)(2l+3)}$ & $-\frac{3}{2(2l-1)(2l+1)}$ & $-\frac{3}{2(2l+1)}$ \\
\end{tabular}
\end{center}
\caption{\label{greentable}Integrals of Green's kernels for a spherical cell.}
\end{table}

\subsection{Explicit solution}
We note that the Green's kernels (\ref{kernels}) are diagonal in the basis of vector spherical harmonics for spherical shape.
Furthermore, we see that the amplitudes of the surface viscosity force $f^v_{3,l,m}$ depend only on $u^c_{3,l,m}.$
This implies that the vector spherical harmonics with first index 3 are completely decoupled from the two other types.
Since $f^e_{3,l,m}=0$ for all $l$ and $m$, we conclude that $u^c_{3,l,m}=0$  for all $l$ and $m$ as well.
Adding the fixed shape condition $(\boldsymbol u^c-\boldsymbol v_s)\boldsymbol\cdot\boldsymbol n=0$ yields that $\boldsymbol u^c-\boldsymbol v_s$ can be written as a surface gradient of some surface potential $U$, as used in the main text.
Or, in spherical harmonics,
\begin{equation}
\label{grads}
u^c_{1,l,m}=\frac{U_{l,m}}{R}\frac{l}{2l+1},\,\,\, u^c_{2,l,m}=\frac{U_{l,m}}{R}\frac{l+1}{2l+1},\,\,\, u^c_{3,l,m}=0 \textrm{ for }l>1.
\end{equation}

Solving the equations (\ref{f}), (\ref{BIEsphere}), and (\ref{grads}) for $U$ and $f^n$ yields
\begin{equation}
\label{lambda}
U_{l,m}=\frac{R(\chi c^\mu-\alpha c^a)_{l,m}}{\lambda_l}.
\end{equation}
\begin{equation}
\label{lambdal}
\lambda_l=\begin{cases}
&3\eta_{in}R+2\eta_{out}R+2\eta_s+2\eta_b\textrm{, for }l=1,\\
&(2l+1)(\eta_{in}+\eta_{out})R+l(l+1)\eta_b+2(l^2+l-1)\eta_s\textrm{, for }l>1.
\end{cases}
\end{equation}
\begin{equation}
\label{vs}
v_s=-\frac{2}{3}\nabla U_1.
\end{equation}
\begin{equation}
\label{fnres}
f^n_{l,m}=-\left[(l+2)\eta_{out}R+(l-1)\eta_{in}R+2(l+2)(l-1)\eta_s\right]\frac{U_{l,m}}{R^2}\textrm{ for }l>1.
\end{equation}
Using the expressions (\ref{lambda}) and (\ref{vs}), the retrograde flow and the swimming velocity can be expressed as a function of the concentration fields.
The time evolution equations of the concentration fields are obtained by substituting $U$ into eqs. (3) and (4) of the main text.
The following equation can be used to reduce all calculations to scalar spherical harmonics
\begin{equation}
\label{Uc}
\boldsymbol\nabla^s\boldsymbol\cdot\left[c(\boldsymbol u^c-\boldsymbol v_s)\right]=\boldsymbol\nabla^s\boldsymbol\cdot(c\boldsymbol\nabla^s U)=\frac{\Delta^s(cU)+c\Delta^sU-U\Delta^sc}{2}.
\end{equation}

\section{Numerical procedure}
The numerical procedure consists in representing $c^a$, $c^\mu$ and $U$ by the amplitudes of the spherical harmonics for all values of $l<l_{max}$ and $|m|\le l$, where $l_{max}$ is the cut-off value.
We take $l_{max}=64$ for such calculations.
We observed that regardless of the initial conditions, the dynamics relaxed to an axisymmetric solution.
We therefore also performed simulations with the shape assumed axisymmetric from the beginning, which is achieved by setting all amplitudes for $m\ne 0$ to zero.
With this assumption, $l_{max}=1024$ was used, which proved to be necessary for strongly polarized cells.
The eqs. (3) and (4) of the main text were solved by an explicit Euler scheme by truncating the harmonic expansion of the advection terms to $l<l_{max}$.
The time step was chosen small enough to avoid the instability due to the stiffness of the diffusion equation (typically $10^{-4}$ in non-dimensional units).
In some cases, a small diffusion of actin (diffusion coefficient  $10^{-3}$ in non-dimensional units) was added to enhance the stability of the actin advection equation.
The steady-state branches in Figs. 2 and 3 of the main text were obtained by solving eqs. (3) and (4) of the main text with $\dot c^a=\dot c^\mu=0$ using Newton's method.

\section{Model for the cell shape}
The calculation of the shape follows the method used in Ref. [28] of the main text.
Spherical shape of a cell in suspension can be physically achieved by a combination of high osmotic pressure $\Delta P$ inside the cell and the inextensibility of the membrane.
Further in this section, we allow the shape of the cell to deviate from a sphere, albeit weakly, taking the leading terms in the small-deformation expansion.
We parametrize the cell shape by a shape function $\rho$
\begin{equation}
\label{sphericalx}
|\boldsymbol x|=R_0\left[1+\rho(\boldsymbol x)\right],
\end{equation}
where $|\boldsymbol x|$ is the distance from the center of the cell to a given point on its boundary.
$\rho_0=0$ to the leading order in deformation because of the conservation of the membrane area.
$\rho_1=0$ because this term corresponds to a translation of the cell to the leading order.
We show below that $\rho$ scales as $\Delta P^{-1},$ which justifies an expansion in powers of $\rho$.
We consider the quasi-spherical limit, taking the leading terms in such expansions.
The function $\rho(\boldsymbol x)$ is expanded in spherical harmonics of $\boldsymbol x$ to be used below
\begin{equation}
\label{rho}
\begin{aligned}
&\rho(\boldsymbol x)=\sum\limits_{l=2}^\infty\rho_l(\boldsymbol x)\\
&\rho_l(\boldsymbol x)=\sum\limits_{m=-l}^l\rho_{l,m}Y_{l,m}(\boldsymbol x).
\end{aligned}
\end{equation}

Assuming the tension of the membrane $\zeta_0$ to be homogeneous (unaffected by the cortex flow) we can write for the tension force $\boldsymbol f^m$
\begin{equation}
\label{fm}
\boldsymbol f^m=-H\zeta_0\boldsymbol n=-\zeta_0H_0\boldsymbol n-(H-H_0)\boldsymbol n\zeta_0,
\end{equation}
where $H$ is the mean curvature of the membrane (sum of the principal curvatures) and $H_0=2/R$ is the value of $H$ for a perfectly spherical cell.
The term $-\zeta_0H_0\boldsymbol n$ in eq. (\ref{fm}) corresponds to an isotropic compression of the fluid inside the cell, which is balanced by the osmotic pressure.
This relates the tension of the membrane $\zeta_0$ to the pressure difference by the Laplace law:
\begin{equation}
\label{Laplace}
\zeta_0= \frac{R \Delta P}{2}.
\end{equation}
The term $(H-H_0)\boldsymbol n\zeta_0$ in eq. (\ref{fm}) corresponds to a position-dependent normal force, which we identify with the Lagrange multiplier $f^n\boldsymbol n$ used to maintain the shape of the cell.
This justifies that $H-H_0$ scales as $\zeta_0^{-1}$ or, equivalently, as $\Delta P^{-1}$ for fixed $f^n$.
Since $f^n$ is governed by the actomyosin dynamics, as follows from eq. (\ref{fnres}), we obtain that the shape of the cell can be indeed made as close to a sphere as necessary by choosing $\Delta P$ large enough.
This shows that all calculations made for perfectly spherical cells remain valid to the leading order in the limit of large $\Delta P$ even if the spherical-shape condition is relaxed.

The mean curvature can be related to the shape function by
\begin{equation}
\label{sphericalH}
H(\boldsymbol x)=\frac{2}{R}+\frac{1}{R}\sum\limits_{l=2}^\infty (l-1)(l+2)\rho_l(\boldsymbol x).
\end{equation}
This yields the final relation between the shape harmonics $\rho_{l,m}$ and the Lagrange multiplier $f^n$:
\begin{equation}
\label{rhores}
\rho_{l,m}=-\frac{2f^n_{l,m}}{(l-1)(l+2)\Delta P}.
\end{equation}

\section{Physical parameters}
We list in table.~\ref{t:valpar} the physical data we have considered to obtain rough estimates of the three non-dimensional parameters entering in the model.

\begin{table}
\scriptsize
\begin{tabular}{lll}
\hline\hline
name & symbol & typical value \\ 
\hline
cortical thickness & $h$ & $10^{-7}$ m \cite{clark2013monitoring,turlier2014furrow}\\
cortical viscosity & $\eta_s$ & $h\times (10^3-10^6)$ Pa m s \cite{turlier2014furrow,bergert2015force}\\
myosin contractility & $ \chi c_0^{\mu}$ &$h\times (10^2-10^3)$ Pa m \cite{bergert2015force,turlier2014furrow} \\
F-actin compressibility & $ \alpha c_0^{a}$  &$h\times 10^3$ Pa.m \cite{hawkins2011spontaneous}\\
myosin diffusion coefficient & $D^{\mu}$ &$10^{-13}-10^{-12}$ $\text{m}^{2}\text{s}^{-1}$ \cite{uehara2010determinants, hawkins2011spontaneous}\\
cell size & $R$ &$10^{-5}$ m  \\
F-actin turnover  & $\beta$ &$ 10^{-2}-10^{-1}$ ~$\text{s}^{-1}$ \cite{hawkins2011spontaneous, fritzsche2013analysis,turlier2014furrow} \\
\hline
characteristic length & $l_0=R$ &$10^{-5}$ m  \\
characteristic time & $t_0=R^2/D^{\mu}$ &  $10^2-10^3$ s \\
characteristic surface stress & $\sigma_0=D^{\mu}\eta_s/R^2$ & $1-10^4$ Pa m \\
\hline
contractility parameter & $\bar{\chi}=\chi c_0^{\mu}/\sigma_0$ & $10^{-2}-10^3$ \\
compressibility parameter & $\bar{\alpha}=\alpha c_0^{a}/\sigma_0$ & $10^{-1}-10^3$ \\
turnover parameter & $\bar{\beta}=\beta t_0$ & $1-10^2$\\
\hline\hline
\end{tabular}
\caption{\small Estimates of material coefficients and  non dimensional parameters definitions.\label{t:valpar}}
\end{table}

\end{document}